\documentclass[prl,preprint,floatfix]{revtex4}
\usepackage{epsfig}

\newcommand{\be}{\begin{equation}}
\newcommand{\ee}{\end{equation}}
\newcommand{\M}{m}

\begin{document}
\vskip -2cm{\rightline{IFT05-35}}

\title{Large $N$ reduction on a twisted torus}
\author{A. Gonz\'alez-Arroyo}
\affiliation{ Instituto de F\'{\i}sica Te\'orica UAM/CSIC \\
Univ. Aut\'onoma de Madrid,\\
Cantoblanco, Madrid 28049, SPAIN \\{\tt antonio.gonzalez-arroyo@uam.es }} 
\author{R. Narayanan}
\affiliation{Department of Physics, Florida International University, Miami,
FL 33199, USA\\{\tt rajamani.narayanan@fiu.edu}}
\author{ H. Neuberger}
\affiliation{
Rutgers University, Department of Physics
and Astronomy,
Piscataway, NJ 08855, USA\\{\tt neuberg@physics.rutgers.edu}
}

\begin{abstract}
We consider $SU(N)$ lattice gauge theory at infinite $N$
defined on a torus with a CP invariant twist. 
Massless fermions are incorporated in an elegant way, 
while keeping them quenched. We present some 
numerical results which suggest that twisting 
can make numerical simulations of planar QCD more efficient. 
\end{abstract}

\maketitle

\section{Introduction}
At an infinite number of colors, QCD on an Euclidean torus of size $l^4$
undergoes a staircase of transitions as $l$ is 
reduced~\cite{knn}. For $l > l_c$,
where in ordinary QCD units $l_c \sim ~1~fermi$, the system is in a phase
where Wilson loop operators of arbitrary size have traces that are
exactly $l$-independent. Thus, one can reduce the number of degrees of freedom
from that of an infinite torus, without any loss of information at leading order
in the $\frac{1}{N}$ expansion. This ought to be of help in getting at planar QCD
using numerical simulation, as reduction holds on the lattice too.  In
practice  this implies  that increasing $N$ reduces the finite size corrections, so 
that a balance between $N$ and the size of the system can be reached which minimizes
the computational effort required to get the planar limit of various physical QCD observables. It
is likely that getting Monte Carlo numbers in the planar limit is cheaper than solving
full QCD with the computer. 

To obtain numbers appropriate for zero temperature infinite volume planar QCD one must
make sure that all simulations are carried out at lattice spacings $a$ and lattice sizes
$L$ which obey $La > l_c$. For sufficiently fine lattices, $L$ turns out to be of order
10. In this paper we aim to reduce this value even further
by making use of  an old idea  due to Gonz\'alez-Arroyo and
Okawa~\cite{twist}. We consider pure $SU(N)$ YM theory on a twisted
torus. At infinite $N$ the large volume phase should 
be independent of the boundary
conditions in as much as it does not depend on the size of the box. 
When the volume is decreased, the theory enters 
a phase where some dependence on boundary conditions is restored.
This phase must be distinct from the corresponding phase in the 
untwisted case, where it corresponds to finite temperature deconfined
planar QCD. Therefore, it is conceivable
that the critical size $l_c^t$ for the twisted box is smaller than $l_c$ and this follows from general arguments. It is also supported
by the quantitative numerical work presented below.

\section{Twisted torus -- pure gauge}

$SU(N)$ gauge fields are objects in $SU(N)/Z(N)$ and therefore the allowed bundles are 
those of $SU(N)/Z(N)$ over the torus~\cite{tonyrev}. 
Some of these bundles cannot be lifted to an $SU(N)$ bundle
and when this happens we'll say that we have a ``twisted torus''. 
In this work we are only interested in 
the CP invariant case where $N$ is assumed to be even. We consider 
a nontrivial $SU(N)/Z(2)$ bundle over
the torus. To ensure that the classical limit is as simple as possible, we further restrict $N$
to be divisible by four. This ensures that there are flat connections in the bundle. 
It is easy to transfer this continuum gauge bundle to the lattice.
If $N$ were
not divisible by four, only by two, the bundle would admit only half integral topological charges and
the minimal action configuration would have nontrivial space-time structure.

We use a single plaquette Wilson lattice action and the gauge group is $SU(N)$, 
where $N=4M$ and $M$ is chosen to be prime. Our choice of  twist can be  induced by choosing
the sign of the lattice coupling $\beta$ to be
negative and taking a symmetrical
hypercubic lattice of volume $L^4$ with $L$ given by an odd integer. As is well known, the
unusual sign of the coupling could be absorbed by a change of lattice gauge variables for even $L$, 
and there is no twist. The same change of variables is inconsistent at the boundaries
when $L$ is odd, where it becomes equivalent to the nontrivial boundary conditions
one would use if one defined the twisted bundle in the continuum by starting from an
open sub-hypercube of $R^4$. 
Another way to see that a negative lattice coupling amounts to twisting by -1 for odd
$L$ is to observe that the Z(2)
flux through any plane becomes  $(-1)^{L^2}$.

The change of variables needed to bring the  twisted action to a negative
coupling Wilson action also affects observables. In particular,  let $C$ 
denote a  closed finite  curve $C$ on an infinite lattice, 
mapped in the natural way to the torus. Associated with $C$ there is 
a sign, $s(C)\equiv (-1)^{p(C)}$, where $p(C)$ is the number of 
plaquettes in a spanning surface with $C$ as 
boundary on the infinite lattice. Let $W_L(C;b,N)$ 
denote a  Wilson loop expectation value associated 
with the curve $C$ in the presence of periodic 
boundary conditions for SU(N) gauge theory at lattice coupling
$\beta=2bN^2$. Then, after the change of variables 
it transforms into $s(C) W_L(C,-b,N)$,
where $W_L(C;-b,N)$ denotes the ordinary Wilson loop 
computed on a
periodic lattice of volume $L^4$ with a 
negative value of the coupling constant.
Inspection of the loop equations led Eguchi 
and Kawai~\cite{ek} to conclude that
in the large N limit $\frac{1}{N} tr W_L (C;b,N)$ 
is $L$ independent. Their proof can
be extended~\cite{twist} 
to twisted boundary conditions. This result can be also
deduced from an analysis of the strong coupling 
expansion~\cite{strong} directly. Thus, 
in the strong coupling region one has:
\begin{eqnarray*}
\lim_{N\to\infty}\frac{1}{N} tr W_L (C;b,N)=\lim_{N\to\infty}\frac{1}{N}
tr W_\infty (C;b,N)= \\
\lim_{N\to\infty}\frac{s(C)}{N} tr
W_L(C,-b,N)=\lim_{N\to\infty}\frac{s(C)}{N} tr W_\infty (C;-b,N)
\end{eqnarray*}
The above equation does not extend all the way
to the continuum limit $|b|\to\infty$~\cite{BHN}, but 
previous work has produced evidence in favor of its
validity beyond the radius of convergence of 
the strong coupling expansion. The basic
idea of this paper is to estimate $\lim_{N\to\infty}
\frac{1}{N} tr W_\infty (C;b,N)$ at some `t Hooft coupling $b>0$ 
by numerically extrapolating
$\frac{s(C)}{N} tr W_L (C;-b,N)$ 
to infinite $N$ at $-b$ and fixed $L$. 

As explained in the introduction, for reduction to work
at a given $b>0$, one needs $L\ge L_c(b)$. 
Based upon past experience with twisting and on 
arguments to be given later we expect that $ L_c(-b) < L_c (b)$.
Our numerical findings indicate  that this is true, opening the way
to more efficient numerical work on planar QCD, employing a CP invariant 
twist. 

The proof of reduction in perturbation theory~\cite{twist,EN,dasrev} 
requires the vacuum structure to be relatively simple. 
For $N$ given by $4M$, where $M$ is prime, 
the minimal action configurations are made up of gauge orbits defined by 
the gauge configuration $U_\mu=\Gamma_\mu\otimes D_\mu$ 
where the $\Gamma_\mu$ are
ordinary 4 by 4 Euclidean Dirac matrices, and the $D_\mu$ 
are diagonal matrices of size $M$.
The moduli space is defined by the $4(M-1)$ angles 
associated with the matrices $D_\mu$.

The system has a global $Z^4(N)$ symmetry, which is 
particularly important at finite
volume. Any one of the vacua labeled by points in the moduli space preserves 
a $Z^4(2)$ subgroup of this symmetry. 
The remainder, $Z^4(M)$, would be restored by 
averaging over the moduli space of angles with flat measure. 
In other words, at
infinite $N$ one can say that the 
eigenvalues of Polyakov loops in all directions
are uniformly distributed over the unit circle.
Thus, uniform averaging would be a correct procedure
if we knew that we are in a phase where the entire $Z^4(N)$ stays 
unbroken even at infinite $N$. 
This average over the moduli space at infinite $N$ effects an extension of the
discrete momentum sums associated with ordinary Feynman diagrams on a torus 
to continuous Feynman integrals on the 
smooth space of crystal momenta normally associated 
with an infinite lattice, with
the angles filling in the momentum gaps typical of a finite spacetime torus. 
As we have seen above, effectively, 
twisting fractionalizes the Brillouin zone into
16 identical hypercubes, facilitating the 
gap-filling role assumed by the remaining
angular parameters. All in all, twisting ``helps'' 
the system to maintain the global
$Z^4(N)$ at $N=\infty$ and this is required for reduction to work.

When $|b|$ is increased too much, one expects 
the global $Z^4(N)$ to break spontaneously
in the large $N$ limit. 
We are not certain of the phase of the theory when $L < L_c(-b)$.
The simplest guess would 
be a breaking of one of the $Z(N)$ factors
down to $Z(2)$. Substantial numerical
work would be needed to determine whether this is correct, 
or if another alternative
takes over. This is an issue we postpone to the future. In this work, we shall 
carry out tests mainly at one particular $b$-coupling. From these results, we 
shall be able to also conclude that in the test cases the 
entire $Z^4(N)$ symmetry group was preserved and that reduction held.

\section{Numerical tests -- pure gauge}

Our first goal is to check whether the twist indeed helps in reducing 
the lattice size $L$ at which we can attain the low temperature symmetric
phase. For that purpose, we  take a lattice spacing $a$ for which we know that reduction works on 
a periodic  lattice only for $L$ larger than a specific $L_c$.
Then, we
try to find out whether a simulation with twist on a smaller size
$L^t < L_c$ torus is able to reproduce the  value of various large  $N$ observables.
Here we make use of the size independence of the results in the large $N$
limit and in this phase. 

In particular, we chose an inverse 
't Hooft coupling, $b$ set to $|b|=0.36$; the corresponding
lattice spacing is $a\approx(2.1~GeV)^{-1}$,
quite typical of current QCD simulations on what is considered a fine lattice. 
Using regular boundary conditions at this lattice spacing would require 
$L\ge 9$. We ran a series of tests which show that a twisted
lattice of size $L^t=5$ at  the same value of $|b|$, is able to remain 
in  the phase where full reduction holds, but staying away from
the lattice strong coupling phase. (The latter phase extends from $|b|=0$ to about
$|b|=0.36$, but at large $N$ the tunneling rate into the strong coupling phase
can be kept so low that one does not need to worry even about going slightly below
$|b|=0.36$.)

Our Wilson loops
were built out of $\tilde U_\mu(x)$ matrices,
rather than the original link matrices $U_\mu (x)$. The  
$\tilde U_\mu(x)$ matrices are defined in term of the  $U_\mu (x)$
by an iterative ``smearing'' procedure~\cite{ape}. Let $\Sigma_{U^{(n)}_\mu (x)}$
denote the ``staple'' associated with the link $U^{(n)}_\mu(x)$ in terms
of the entire set of $U^{(n)}_\nu(y)$ matrices. 
One step in the iteration
takes one from a set $U^{(n)}_\mu (x)$ to a set $U^{(n+1)}_\mu (x)$,
by the following equation:
\begin{eqnarray*}
X^{(n+1)}_\mu (x)\equiv\alpha U^{(n)}_\mu (x)+\frac{1-\alpha}{6}
\Sigma_{U^{(n)}_\mu (x)}\nonumber\\
U^{(n+1)}_\mu (x)=X^{(n+1)}_\mu (x)
\frac{1}{\sqrt{[X^{(n+1)}_\mu (x)]^\dagger X^{(n+1)}_\mu (x)}}
\end{eqnarray*}
We chose $\alpha=0.45$ in the untwisted case  and iterated twice:
\begin{eqnarray*}
\tilde U_\mu(x)=U^{(2)}_\mu (x)
\end{eqnarray*}
Given the change of variables mentioned previously this changes to 
$\alpha=-0.45$ in the twisted
case.  
Smearing has the effect of removing some of the ultraviolet fluctuations and produces
more meaningful numbers for our comparison. Also, the test is made more stringent
by including smearing because smearing is a relatively complicated operation
involving longer  loops and, although it should survive twisting theoretically, 
numerical effects might have marred the equivalence to the untwisted case.

\begin{figure}
\epsfxsize = 1.0\textwidth
\centerline{\epsfbox{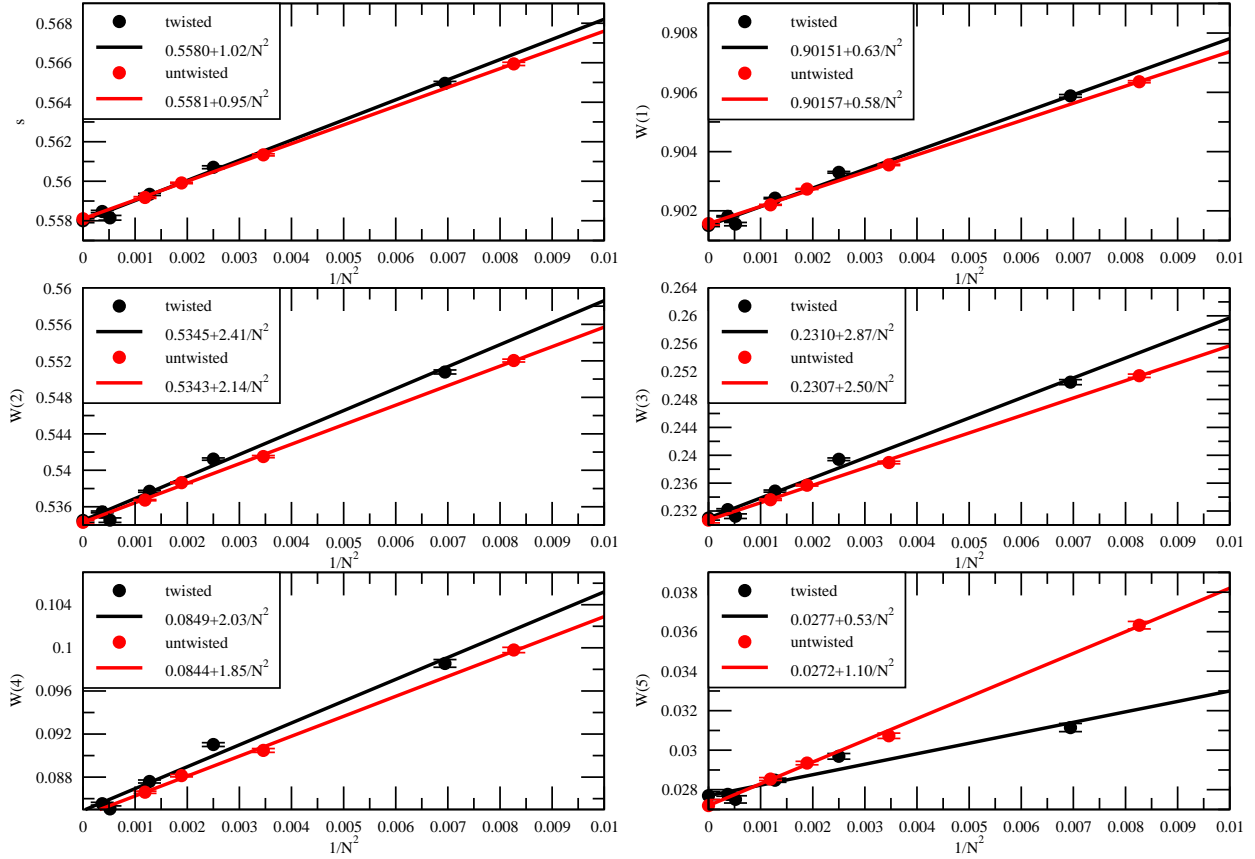}}
\caption{Results for the untwisted case on a $9^4$ lattice and for the 
twisted case on a $5^4$ lattice at coupling $|b|\equiv\frac{|\beta|}{2N^2}=0.36$
}
\label{cumplot}
\end{figure}

In Table~\ref{tab1}, 
we compare untwisted results obtained from the extrapolation to infinite
$N$ of a sequence of $b=0.36$ $9^4$ lattices with $N=11,17,23,29$ to results obtained
from the extrapolation to infinite $N$ of a sequence of $b=-0.36$ $5^4$ 
lattices with $N=12,20,28,44,52$. These were linear extrapolations
in $\frac{1}{N^2}$ and are shown in Figure~\ref{cumplot}.
We compare the action density (raw plaquette average), and smeared $n$ by $n$ Wilson loops
for $n=1,2,3,4,5$. The agreement is within error bars, but the large $N$ extrapolation
works better in the untwisted case; nevertheless, the coefficients of the $\frac{1}{N^2}$
correction come out quite close in the twisted and untwisted cases, except in the twisted case
of the $5\times 5$ Wilson loop, where the linear fit in $\frac{1}{N^2}$ does not work well.
Theoretically, one expects that the $\frac{1}{N^2}$ correction depend on the shape of the box,
and since the twisted box is of size $5^4$, this is a natural place to see some larger
corrections. 
There is no question that the twisted simulations took
less computer time, but a more quantitative comparison of efficiency needs to
take into account the errors. The errors in
Table~\ref{tab1} were estimated without taking correlations into
account because we did varied kinds or runs. The quoted 
errors should be viewed just as rough indicators. 
We don't have enough information for a quantitative efficiency estimate, but there is
little doubt that it pays to twist. 

\begin{table}
\caption{\label{tab1}Action density $s$ and smeared  $n\times n$ Wilson loops $W(n)$ 
on twisted and untwisted lattices.} 
\begin{ruledtabular}
\begin{tabular}{lcr}
Operator & untwisted  &twisted\\
\hline
$s$ &0.5581(1)  &0.5580(1)\\
$W(1)$ &0.90157(5)  &0.90151(4)\\
$W(2)$ &0.5343(2)&0.5345(2)\\
$W(3)$ &0.2307(4)&0.2310(3)\\
$W(4)$ &0.0844(4)&0.0849(3)\\
$W(5)$ &0.0272(3) &0.0277(2)\\
\end{tabular}
\end{ruledtabular}
\end{table}

\section{Twisted torus -- Fermions}

Perhaps the main numerical advantage of conventional planar QCD over real QCD
is that fermions are quenched. This poses a problem on the
twisted torus, as the
fermions are in the fundamental of $SU(N)$ and no longer transform under just
$SU(N)/Z(N)$. Another set of old tricks~\cite{venez} can be brought to bear on this problem. First,
enlarge the gauge group to $SU(N)\otimes SU(K)$. Next recall that with the choice
of twists made here, only $SU(N)/Z(2)$ was exploited. Consider now the true group
to be $[SU(N)/Z(2)]\otimes [SU(K)/Z(2)]$ and make the fermions transform under
the latter as a bi-fundamental, canceling the $Z(2)$ twists between
the two gauge groups.  Now everything is in order and all we need to do
is to take $N$ and $K$ divisible by 4. We choose $K=4$ and $N=4M$ with prime $M$,
as before. We also wish to get rid of the gauge fields associated with the $SU(K)$
factor. We take its lattice $|b|$ coupling to infinity
and freeze out those degrees of
freedom. Fermion observables that are 
singlets 
under the full $SU(N)\otimes SU(K)$ fermions
are still quenched in the large $N$ limit. We expect there to
be no difference between the twisted theory and the regular one at $N=\infty$, if we take 
four non-interacting flavors in the
periodic case. For quenched fermions, the number of flavors is immaterial
in the regular case, as the Dirac operator block decomposes over flavors. In the
twisted case the flavors are coupled and this increases the cost of the fermion
simulation relatively to the untwisted case. It is quite possible that this increase along with the need to go to larger $N$
is outweighed by the smaller $L$ one needs -- only experimentation can determine the
cost effectiveness of twisted torus for fermion simulations in the planar limit. 

When the fermion twisting trick is taken to the lattice more checks are needed.
First of all, we certainly want to preserve the hard won option of maintaining
exact global chiral symmetry at finite lattice spacing~\cite{overlap}. In the untwisted case
we know how to do this for each fermion flavor independently. Now we need to
make sure that the coupling of the flavors in the specific way associated with 
twisting can be 
taken to the lattice. We cannot first put each flavor on the lattice and then
couple them, because we would loose explicit 
lattice translational invariance. So, we must first couple
the flavors and then carry out the overlap construction, as it is obvious that
the twisting procedure meshes well with the sparse Wilson lattice fermion action. 

In the untwisted case, the bilinear fermionic action for one flavor 
is described by the massless overlap
Dirac operator $D_o$~\cite{overlap}:
\begin{eqnarray*}
&D_o = \frac{1+V}{2}\nonumber\\
&V^{-1}=V^\dagger=\gamma_5 V \gamma_5
={\rm sign}(H_w (\M))\gamma_5
\end{eqnarray*}
$H_w (m)$ is the Wilson Dirac operator at mass $\M$, which we shall choose
as $\M=-1.5$. 
\begin{eqnarray*}
H_w (M)= \gamma_5 \left [ 
\M+4 -\sum_\mu \left ( \frac{1-\gamma_\mu}{2} T_\mu +\frac {1+\gamma_\mu}{2} 
T_\mu^\dagger \right ) \right ]
\end{eqnarray*}
The $T_\mu$ matrices are the lattice generators of parallel transport and
depend parametrically and analytically on the lattice links $U_\mu(x)$
which are $SU(N)$ matrices at site $x$ associated with the link connecting
site $x$ to site $x+\hat\mu$, where $\hat\mu$ is a unit vector in the
positive $\mu$-direction. 

The internal fermion-line propagator,
$\frac{2}{1+V}$ is not needed at infinite $N$, as the fermions
are quenched at leading order in $\frac{1}{N}$. For fermion lines continuing
external fermion sources we are allowed to use a slightly different
quark propagator~\cite{overprop,EHN} defined by:
\begin{eqnarray*}
\frac{1}{A} =\frac{1-V}{1+V}
\end{eqnarray*}
$A=-A^\dagger$ and anticommutes with $\gamma_5$. The spectrum of $A$ 
is unbounded, but is determined by the spectrum
of $V$ which is restricted to the unit circle. One should
think of $A$ as dimensionless, and of $|\M|$ as providing
the needed dimension. Up to a dimensional unit, 
$A$ should be thought of as a lattice realization of 
the continuum massless Dirac operator, $D$:
\begin{eqnarray*}
2|\M| A \leftrightarrow D=\gamma_\mu \partial_\mu + .....
\end{eqnarray*}

Twisting involves the $T_\mu$ operators, which would act now on two
indices, color and flavor. There is also the notational inconvenience since
we have to deal with three kinds of gamma matrices: the Lorentz $\gamma_\mu$ from above,
the $\Gamma_\mu$ of color space that enter the classical vacua, and now, in
addition, $\hat\gamma_\mu$ acting on flavor. To include the twist for 
fermions, the new $T_\mu$ operators are extended by:
\begin{eqnarray*}
T_\mu \rightarrow T_\mu\otimes{\hat\gamma_\mu}
\end{eqnarray*}
As a result, the rank of $D_o$ increases four fold. 
One still has global flavor singlet chiral symmetry, but no flavor-non-singlet symmetry.
Indeed, flavor is not a symmetry, as it implements the twist. It remains
to check that in perturbation theory one effectively has four species of Dirac fermions.

To see this, we go to one of the vacua where all diagonal matrices 
$D_\mu$ are unity and can be suppressed. (We already know that when 
they are not unity they effectively induce some small amount of 
gap-filling momentum into the fermion lines.) Once the $D_\mu$ are ignored, 
the $T_\mu$ matrices get replaced by 
$\Gamma_\mu\otimes{\hat\gamma_\mu }$ for each $\mu$. One 
needs now to diagonalize $H_w$. Specifically, the focus is 
on the gauge dependent Wilson mass term, defined as 
$4-\sum_\mu (\pm ) \Gamma_\mu\otimes {\hat\gamma_\mu}$. There, with
all momenta zero and any Dirac index, one finds sixteen states with 
eigenvalues 8,6,4,2,0 and respective degeneracies of 1,4,6,4,1. 
As indicated by the $\pm$ signs, one also needs to take into account all other 15 momenta 
where some subset of zero momentum components get replaced by 
$\pi$. The set of eigenvalues and degeneracies stays the same for 
each one of the sixteen momenta. With our choice of the parameter 
$\M$, only the sixteen states of zero eigenvalue will produce a 
massless Dirac fermion. So, in total we obtain $16M$ massless 
Dirac fermions, including all flavors and colors. Since the number 
of colors is $N=4M$ we have $4N$ fermions, exactly as expected in 
the continuum: four flavors of colored fundamental multiplets. 

The way this worked out is quite remarkable. Similarly to staggered fermions, the split
of the Brillouin zone into sixteen components contributes one species for each compartment.
However, unlike in the case of staggered fermions, ordinary Dirac indices are not
being mixed in and the global chiral symmetry is the ordinary continuous group we know from 
continuum. Only flavor is scrambled up, but it had to be, because if the fermionic
action fully factorized, we would have concluded that one can define a single fermion
flavor on our twisted bundle, something that is geometrically impossible in the continuum.
However, in the planar limit there is an equivalence of the 
flavor and gauge singlet Green functions of the twisted theory 
to an untwisted theory in which the four flavors are decoupled. 
This does not really mean that the flavors are decoupled in the twisted theory; rather, 
so long as one is restricted to only considering flavor singlets, the fermion
flavors act as if they were decoupled. 

Given the rather intricate nature of this mechanism, a numerical 
check is highly desirable; as we shall see, it works very well. 
Before turning over to numerical results, we wish to point out that had we
been interested in dynamical fermions, the case of fermions in the double indexed antisymmetric
representation of $SU(N)$ is an excellent candidate theoretically, as it is unaffected by
our $Z(2)$ twist. In this case, no extra flavors are needed. Thus, for projects trying
to get at the planar limit of  
supersymmetric QCD~\cite{susy}, our $Z(2)$ twist poses no fermionic problems.

\section{Numerical tests -- Fermions}

Previous work employing an untwisted action at infinite $N$ in the physical phase, 
showed how the use of random matrix theory~\cite{rmt} to calculate the 
fermion condensate and establish spontaneous chiral symmetry breaking
on the lattice. 
In the twisted case, at infinite $N$, we can do the same. We wish to check that 
after proper rescaling we find a condensate identical to the condensate we found using
the regular method, just multiplied by four on account of the four flavors~\cite{bpsipsi}.

\begin{figure}
\epsfxsize = 1.0\textwidth
\centerline{\epsfbox{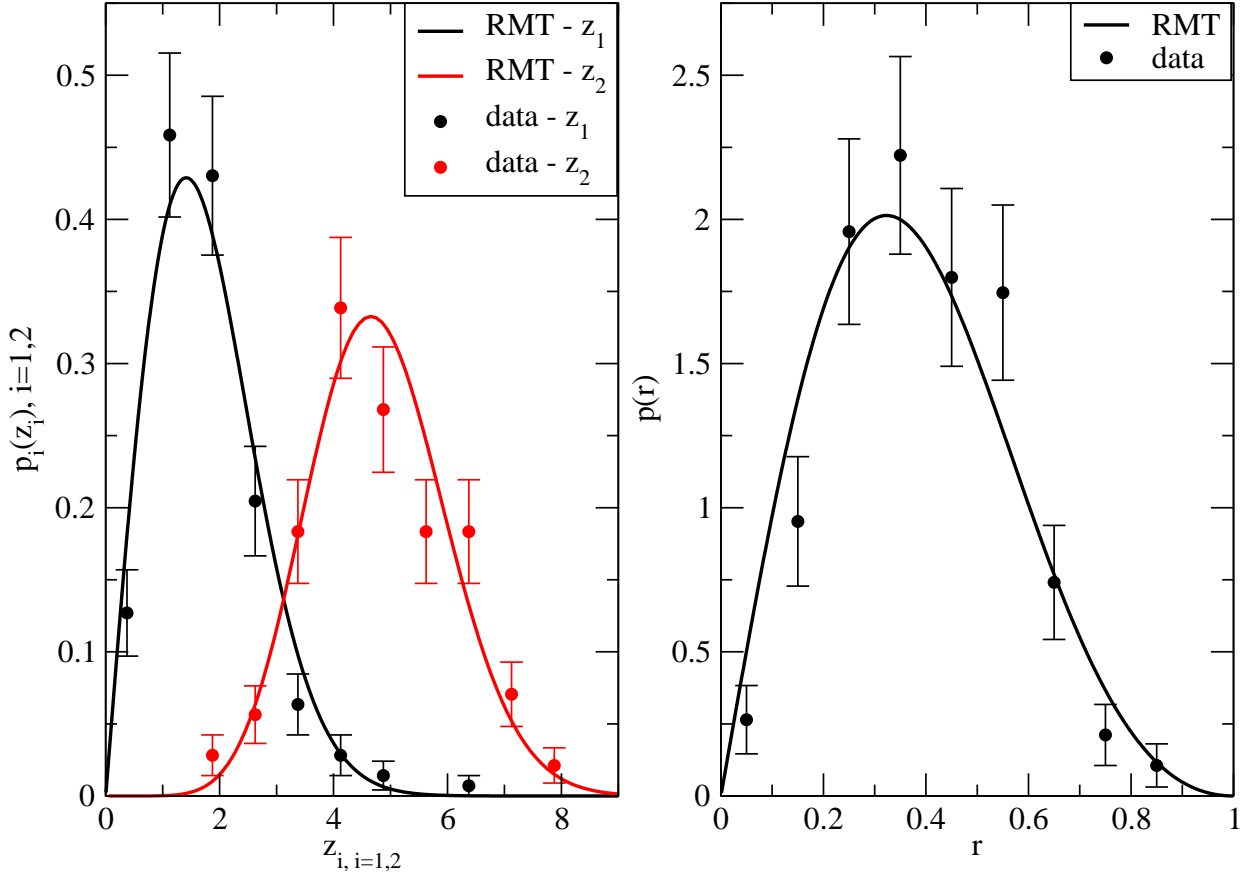}}
\caption{Distributions associated with overlap Dirac 
operator eigenvalues for the 
twisted case on a $5^4$ lattice at coupling $|b|\equiv\frac{|\beta|}{2N^2}=0.35$
}
\label{rmt35}
\end{figure}

In the untwisted case we gathered a large amount of data at $b=0.35$, which
is a coarser lattice. To facilitate comparison of the bare quantities directly
we now focus on the twisted case with at $b=-0.35$ on a $5^4$ lattice
at $N=44$. We first establish that the two lowest eigenvalues of $\sqrt{-A^2}$ 
indeed have a ratio $r$ distributed by the parameter free prediction $p(r)$ 
of the Shuryak-Verbaarschot model. This is shown in the 
right panel of Figure~\ref{rmt35}. 
Next, consistent estimates for the
condensate can be extracted using the two smallest eigenvalues independently:
after scaling each by a fitted number their distributions are predicted
to be given by universal functions, $p_i (z_i),~i=1,2$ where $z_i$ is the
rescaled value of the $i$-th eigenvalue. The two fits are consistent with
each other and are compared to the data in the left panel of Figure~\ref{rmt35}.
We have generated 189 gauge configurations; from the lowest eigenvalue we obtain
$\frac{1}{N}\langle\bar\psi\psi\rangle = (0.140(2))^3$ and from the second lowest
we get $\frac{1}{N}\langle\bar\psi\psi\rangle = (0.143(1))^3$. In the untwisted
case~\cite{bpsipsi} the result at $b=0.35$ 
was $\frac{1}{N}\langle\bar\psi\psi\rangle = (0.142(6))^3$. 

We have also gone to the finer lattice spacing corresponding to $b=-0.36$ in the
twisted case keeping $N=44$ and $L=5$. Here we accumulated 480 configurations, as the decrease in lattice
spacing increases the numerical values of the bare eigenvalues, speeding up the
algorithm that calculates them. Now we see a small but clear deviation of the
ratio distribution away from the universal curve, showing preference for higher
ratios, as typical in these cases, where eigenvalue repulsion has not yet become
fully active. Therefore, one needs to increase the $N$ or $L$
in order to reach agreement with random matrix theory.
This does not rule out that the lowest eigenvalue is already 
correctly distributed and indeed the match to theory with a fitted condensate looks
fine. This gives us  $\frac{1}{N}\langle\bar\psi\psi\rangle = (0.106(1))^3$.
We do not have untwisted data at $b=0.36$, but extrapolating from the data at the
smaller $b$ couplings we do have, we can roughly estimate that 
$\frac{1}{N}\langle\bar\psi\psi\rangle = (0.106(4))^3$ would be the result
at $b=0.36$. This looks good in comparison, but we must keep in mind that the
second eigenvalue in the twisted case also appears to obey its universal prediction
after rescaling, but now we get $\frac{1}{N}\langle\bar\psi\psi\rangle = (0.1102(4))^3$.
Therefore, the second eigenvalue has not yet converged to its random matrix
distribution, and we cannot be sure that the first already has, although the indication
is that it did.

\section{Three dimensions}

Very similar constructions hold in there dimensions, with the obvious difference that
one has no topological charge to worry about and one can take $N=2M$ with $M$ prime.
Brief tests we have carried out in three dimensions also indicate that
twisting 
allows one to deal with fine lattices at smaller lattice volumes than in the regular
approach. 

\section{Summary and Discussion}

The outstanding question is to clearly 
characterize the phase that the twisted system decays into
when $l$ decreases to just below $l_c^t$ 
and numerically check that this phase survives the
continuum limit. Until this is done, one 
cannot claim to have an understanding of
twisted simulations at the same level as 
of untwisted ones. Twisted simulations 
hold the promise of substantial 
savings in computer time due to the ability to
work at even smaller volumes than when 
using conventional periodic boundary conditions on the torus.

For our twisted simulations we used our 
regular untwisted code, and simply used a negative
$b$ and a negative $\alpha$, leaving everything else the same. 
Our regular code was tuned
for the untwisted case, but seemed to 
perform reasonably also on the twisted case. 
Further work is needed to tune a code specifically for the 
twisted case, specifically for weaker couplings than the ones
used in untwisted numerical simulations. 

\section{Acknowledgments}

A.~G-A acknowledges financial support from Spanish Ministry of Education 
through  grant FPA2003-03801. He also wants to thank H. Neuberger and the NHETC at
Rutgers University for the invitation and wonderful hosting during the
early stages of this work.
R. N. acknowledges partial support by the NSF under
grant number PHY-0300065 and also partial support from Jefferson 
Lab. The Thomas Jefferson National Accelerator Facility
(Jefferson Lab) is operated by the Southeastern Universities Research
Association (SURA) under DOE contract DE-AC05-84ER40150.
H. N. acknowledges partial support 
by the DOE under grant number 
DE-FG02-01ER41165 at Rutgers University.

\end{document}